\documentclass[
    ,final            
  ]
  {aipproc}

\layoutstyle{6x9}

\begin{document}

\title{Linear and Weakly Nonlinear Analysis of the Magneto-Rotational-Instability in Thin Keplerian Discs}


\author{E. Liverts}{
  address={Department of Mechanical Engineering\\ Ben-Gurion University of the Negev\\P.O. Box 653, Beer-Sheva 84105\\Israel}
}

\author{Yu. Shtemler}{
address={Department of Mechanical Engineering\\ Ben-Gurion University of the Negev\\P.O. Box 653, Beer-Sheva 84105\\Israel}
}

\author{M. Mond}{
address={Department of Mechanical Engineering\\ Ben-Gurion University of the Negev\\P.O. Box 653, Beer-Sheva 84105\\Israel}
}

\begin{abstract}
 The linear instability of thin, vertically-isothermal Keplerian discs, under the influence of axial magnetic field is investigated. Solutions of the stability problem are found explicitly by asymptotic expansions in the small aspect ratio of the disc. It is shown that the perturbations are decoupled into in-plane and vertical modes. Exact expressions for the growth rates as well as the number of unstable modes are derived. Those are the discrete counterpart of the continuous infinite homogeneous cylinder magnetorotational (MRI) spectrum. In addition, a weakly nonlinear analysis of the MRI is performed. It is shown that near the instability threshold the latter is saturated by the stable magnetoacoustic modes.
\end{abstract}

\maketitle


\section{Introduction}

  The stability properties of Keplerian disks have been a focus of intensive investigation of theoretical astrophysicists over the last decades, pertaining to the problem of angular momentum transfer in accretion disks.  Traditionally, results of the analytical study of magneto-rotational instability (MRI) in an infinitely long cylinder (\cite{Velikhov 1959};  \cite{Chandrasekhar 1960}) have been  adopted for thin disks in order to derive criteria for the spectral stability under various conditions (\cite{Balbus and Hawley 1991}). In addition, it has recently been shown that if the thin disk geometry is taken into account both the growth rates as well as the number of unstable MRI modes are greatly reduced and are decreasing functions of the disk thickness (\cite{Coppi and Keyes  2003};  \cite{Liverts and Mond  2009}; \cite{Shtemler et al.  2011}). The aim of the current work is twofold:1. to provide a complete description of the stable as well as of the unstable spectrum of thin disks under the influence of an axial magnetic field, and 2. to carry out a weakly nonlinear analysis of the MRI in order to investigate its saturation mechanism.

\section{The physical model of thin Keplerian disks}
The stability of radially as well as axially stratified rotating plasma in thin vertically isothermal discs threaded by a magnetic field is considered. Viscosity, electrical resistivity, and radiation effects are ignored.
\subsection{Governing equations}
As a first step, all physical variables are transformed to non-dimensional quantities by using the following characteristic values  (\cite{Shtemler et al. 2011}; \cite{Shtemler et al. 2009}):
$$
t_*=\frac{1}{\Omega_*},\,\,V_*=\frac{r_*}{t_*},\,\,
L_*=V_* t_*,\,\,
 m_*= m_i,\
n_*=n_i,\,\,\,
\,\,\,\,\,\,\,\,\,\,\,\,\,\,\,\,\,\,\,\,\,\,\,\,\,\,\,\,\,\,\,\,\,\,\,\,\,\,\,\,\,\,\,\,\,\,\,\,\,\,\,\,\,\,\,\,\,\,\,\,
\,\,\,\,\,\,\,\,\,\,\,\,\,\,\,\,\,\,\,\,\,\,\,\,\,\,\,\,\,\,\,\,\,\,\,\,\,\,\,\,\,\,\,\,\,\,\,\,\,\,\,\,\,\,\,\,\,\,\,\,
\,\,\,\,\,\,\,\,\,\,\,\,\,\,\,\,\,\,\,\,\,\,\,\,\,\,\,\,\,\,\,\,\,\,\,\,\,\,\,\,\,\,\,\,\,\,\,\,\,\,\,\,\,\,\,\,\,\,\,\,
$$
\begin{equation}
\Phi_*={V_*}^2,\,\,
c_{S*}=\sqrt{T_*/m_*},\,\,\,P_*=m_* n_* c_{S*}^{\,\,2},\,\,
j_*=\frac{c}{4\pi}\frac{B_*}{r_*},\,\,
E_*=\frac{V_* B_*}{c}.\,\,
\,\,\,\,\,\,\,\,\,\,\,\,\,\,\,\,\,\,\,\,\,\,\,\,\,\,\,\,\,\,
\label{1}
\end{equation}
Here $\Omega_*=(GM_c/r_*^3)^{1/2}$  is the Keplerian angular
velocity of the fluid at the characteristic radius $r_*$ that
belongs to the Keplerian portion of the disc;
$G$  is the gravitational constant; $M_c$ is the total mass of the
central object; $c$ is the speed of light; $\Phi_*$  is the characteristic value of the
gravitational potential; the characteristic
mass and number density equal to the ion mass and number density, $m_*=m_i$  and $n_*=n_i$.
The characteristic values of the electric current density and electric field, $j_*$
and $E_*$   have been chosen consistently with Maxwell's equations; $c_{S*}$ is the characteristic
sound velocity; $T_*=T(r_*)$  is the characteristic temperature;
the characteristic magnetic field is
 $B_*=B_z(r_*)$.
The dimensional equilibrium temperature  $T(r)$  and  $B_z(r)$, are free functions.

The resulting dimensionless dynamical equations for vertically isothermal discs are:
\begin{equation}
\frac{ D \bf {V} }{Dt} = -\frac{ 1}{M_S^2}\frac{\nabla P}{n}-  \nabla \Phi+\frac{1}{\beta M^2_S} \frac{\bf{j}
\times {\bf{B}}}{n},\\
\label{2}
\end{equation}
\begin{equation}
\frac{\partial n} {\partial t}
               + \nabla \cdot(n {\bf{V}} ) =0,\\
\label{3}
\end{equation}
\begin{equation}
\frac{\partial {\bf{B} }} {\partial t}+
 \nabla \times
 {\bf{E}}=0,\,\nabla \cdot {\bf{B}}=0,
\label{4}
\end{equation}
\begin{equation}
 {\bf {E} }=  -  {\bf{V}} \times {\bf{B}},
\label{5}
\end{equation}
\begin{equation}
P =nT.
\label{6}
\end{equation}
Here $\nabla P=\bar{c}^2_S\nabla n$  for vertically isothermal discs, and the dimensionless equilibrium sound
speed is given by $\bar{c}^2_S =\partial P/\partial n\equiv T(r)$. Standard cylindrical coordinates
$\{r,\theta,z\}$   are adopted throughout the paper with the associated unit vectors
$\{\bf{i}_r,\bf{i}_\theta,\bf{i}_z\}$; $\bf{V}$  is the plasma velocity; $t$ is time; $D/Dt=\partial/\partial
t+(\bf{V}\cdot\nabla)$  is the material derivative; $\Phi(r,z)=-(r^2+z^2)^{-1/2}$  is the
 gravitational potential due to the central object; $\bf{B}$
   is the magnetic field, $\bf{j}=\nabla\times\bf{B}$  is the current density; $\bf{E}$
  is the electric field; $P=P_e+P_i$  is the total plasma pressure; $P_l=n_lT_l$  are the partial species
  pressures ($l=e,i$); $T=T_e=T_i$  is the plasma temperature; subscripts $e$ and $i$ denote electrons
  and ions, respectively. Note that a preferred direction is tacitly defined here, namely, the
  positive direction of the $z$ axis is chosen according to positive Keplerian rotation. The dimensionless
  coefficients $M_S$  and $\beta$  are the Mach number and the characteristic plasma beta, respectively:
\begin{equation}
M_S=\frac{V_*} {c_{S*}},\,\, \beta=\frac{P_*} {B_{*}^2}.
\label{7}
\end{equation}
Zero conditions at infinity for the in-plane magnetic field and the number density are adopted, namely:
\begin{equation}
B_r=0,\,\,\, B_\theta=0,\,\,\,\,n=0 \,\,\,\, \mbox{for} \,\,\,z=\pm \infty.
\label{8}
\end{equation}
To simplify the further treatment of Maxwell's equations, both the hydro-magnetic basic configuration and perturbations are assumed to be axisymmetric.

A common property of thin Keplerian discs is their highly
compressible motion with large Mach numbers (\cite{Frank et al.
2002}). Furthermore, the characteristic effective semi-thickness of the equilibrium disc $H_* = H(r_*)\,\, (H =
H(r)$ is the local semi-thickness) is defined so that the disc aspect ratio $\epsilon$  equals the inverse
Mach number:
\begin{equation}
\frac{1}{M_S}=\epsilon=\frac{H_*}{r_*}
\ll 1.
\label{9}
\end{equation}
Thus, the thin disc approximation means
\begin{equation}
\frac{1}{M_S}=\sqrt{\frac{r_*T_*}{GM_c}}=\epsilon\ll 1,\,\,
\label{10}
\end{equation}
where
$$
M_S=\frac{V_*}{c_{S*}},\,\,\,V_*=r_*\Omega_*,\,\,\*\Omega_*=\sqrt{\frac{GM_c}{r_*^3}},
\,\,\, c_{S*}=\sqrt{T_*},\,\,\, H_*= \frac{c_{S*}}{\Omega_*} \,.
\,\,\,\,\,\,\,\,\,\,\,\,\,\,\,\,\,\,\,\,\,\,\,\,\,\,\,\,\,\,\,\,\,\,\,\,\,\,\,\,\,\,\,\,\,\,\,\,\,\,
\,\,\,\,\,\,\,\,\,\,\,\,\,\,\,\,\,\,\,\,\,\,\,\,\,\,\,\,\,\,\,\,\,\,\,\,\,\,\,\,\,\,\,\,\,\,\,\,\,\,
\,\,\,\,\,\,\,\,\,\,\,\,\,\,\,\,\,\,\,\,\,\,\,\,\,\,\,\,\,\,\,\,\,\,\,\,\,\,\,\,\,\,\,\,\,\,\,\,\,\,
$$
The smallness of $\epsilon$  means that dimensionless axial coordinate is also small, i.e. $z/r_*\sim
\epsilon\,\,({\mid} z{\mid}  ^{<}_\sim H_*)$, and consequently the following rescaled quantities may be introduced in
order to
further apply the asymptotic expansions in $\epsilon $  [(similar to \cite{Shtemler et al. 2011}; \cite{Shtemler et al. 2009}; \cite{Shtemler et al. 2010}):
\begin{equation}
\zeta=\frac{z}{\epsilon}\sim\epsilon^0,\,\,\,\bar{H}(r) =\frac{H(r)}{\epsilon}\sim\epsilon^0,
\label{11}
\end{equation}
where $\bar{H}(r)=\bar{c}_S(r)/\bar{\Omega}(r)$  is the scaled semi-thickness of the disc.
\subsection{Equilibrium configurations}

We start by deriving the steady state solution. It is first noted that the asymptotic expansion
for the time-independent gravitational potential is given by:
\begin{equation}
\Phi(r,\zeta)=\bar{\Phi}(r)+\epsilon^2\bar{\phi}(r,\zeta),\,\,\,
\bar{\Phi} (r)=-\frac{1}{r},\,\,\, \bar{\phi}(r,\zeta)
=\frac{1}{2}\zeta^2\bar{\Omega}^2(r)+O(\epsilon^2),\,\,\, r>1>>\epsilon.
\label{12}
\end{equation}
Substituting (\ref{12}) into Eqs. (\ref{2})-(\ref{6}) and  setting the partial derivatives with respect to time to zero yield
to leading order in $\epsilon$
 \begin{equation}
\frac{\bar{V}_\theta^2}{r}=\frac{d\bar{\Phi}(r)}{dr},\,\,\,\,
\frac{\bar{c}_S^2(r)}{\bar{n}}\frac{\partial \bar{n}}{\partial \zeta}=
-\frac{\partial \bar{\phi}(r,\zeta)}{ \partial
\zeta}.
\label{13}
\end{equation}
Thus, the  velocity as well as the number density are given by:
\begin{equation}
V_r=o(\epsilon^2),\,\,\, V_\theta =\epsilon ^0\bar{V}_\theta(r)+O(\epsilon ^2),\,\,\,
V_z=o(\epsilon ^2),\,\,\, n\cong\epsilon ^0 \bar{n}\equiv \epsilon ^0\bar{N}(r)\bar{\nu}(\eta),
\label{14}
\end{equation}
where $o(\epsilon)\ll \epsilon, \, O(\epsilon)\sim \epsilon$, and
\begin{equation}
\bar{V}_\theta(r)=r\bar{\Omega} (r),\,\,\, \bar{\Omega} (r)=r^{-3/2},\,\,\,
\bar{\nu}( \eta )=\exp(-\eta ^2/2), \ \ \ \eta=\zeta/\bar{H}(r).
\label{15}
\end{equation}

There are two solutions of eqs. (\ref{13}) - (\ref{15}) that describe two quite different equilibrium magnetic configurations, namely one with comparable magnitudes of the axial and toroidal components of the magnetic field,
 and the other with dominant toroidal component. The two equilibria are distinguished by different scaling of the physical variables with   $\epsilon$. A detailed description of those two equilibria may be found in \cite{Shtemler et al. 2011}. Here however, we focus on the equilibrium that is characterized by comparable magnitudes of the axial and toroidal components of the magnetic field, and in particular will consider the case of a pure axial equilibrium magnetic field. As is shown in \cite{Shtemler et al. 2011}, the results may be easily extended to the more general case of a comparable toroidal component. All equilibrium variables are written in the leading order in  $\epsilon$, and depend on the radial variable only. The exceptions are the number density and the pressure that depend on the axial coordinate in a self-similar manner with radius-dependent amplitude. The axial magnetic field as well as the disc thickness and the amplitude factor, $\bar{N}(r)$, in the number density are arbitrary functions of the radial variable.

To start the equilibrium description it is first assumed that the axial component of the magnetic field is of order  $\epsilon^0$.
That assumption together with relations (\ref{12})-(\ref{15}), determine the order in $\epsilon$  of the rest of the physical variables. The result is given by:
\begin{eqnarray}
B_r = o(\epsilon^2),\,\,\, B_\theta \cong\epsilon^1 \bar{B}_\theta(r),\,\,\, B_z \cong\epsilon ^0 \bar{B}_z(r),\\
j_r = o(\epsilon^2),\,\,\,
j_ \theta \cong\epsilon ^0\bar{j}_ \theta(r)\equiv -\epsilon ^0 \frac{d \bar{B}_z}{d r},\,\,\,
j_z \cong\epsilon ^0 \bar{j}_z \equiv \epsilon ^0 \frac{1}{r} \frac{d (r\bar{B}_\theta)}{d r}.\,\,\,
\label{16}
\end{eqnarray}

It is noted finally that to lowest order in $\epsilon$  the magnetic field configurations under consideration do not influence the steady-state properties of the disk. As will be seen in the following sections, this situation changes dramatically when small perturbations are considered.

\subsection{Perturbed thin discs}

In general for the unsteady nonlinear case the dependent variables are scaled in $\epsilon$  in the following way:
\begin{equation}
f(r,\zeta,t)=\epsilon^{\bar{S}}\bar{f}(r,\zeta)+\epsilon^{S'} f '(r,\zeta,t).
\label{18}
\end{equation}
Here $f(r,\zeta,t)$  stands for any dependent variable, the bar and the prime denote equilibrium and
 perturbed variables; each perturbed variable is characterized by some power
 $S'$, as is summarized in Table 1.



\section{DYNAMICAL EQUATIONS FOR THE PERTURBED DISCS}
As stated above, a detailed stability study is carried out for zero toroidal magnetic field. As shown in \cite{Shtemler et al. 2011} the results of such analysis may be extended to the case of comparable poloidal and toroidal equilibrium components.

\subsection{ The reduced nonlinear equations}
 The perturbations are subject to the boundary conditions for the in-plane magnetic field and number density as follows  from
 (8)
\begin{equation}
B_r'=0,\,\,\,B_\theta'=0,\,\,\,n'=0\,\,\,\,\, \mbox{for} \,\,\,\,\,\,\zeta=\pm \infty.
\label{25}
\end{equation}
%
Both poloidal and toroidal components of the perturbed magnetic field, $B_r'$  and   $B_z'$ , are expressed through the magnetic flux function,   $\Psi'$, which renders the magnetic field divergent free,  $\nabla \cdot {\bf{B}}=0$.

 \begin{table}
 \centering
 \begin{minipage}{140mm}
\caption{The $\epsilon$-scaling of the perturbed variables. }
\begin{tabular}{@{}ccccccccccc@{}}
\hline
$f=\epsilon^{\bar{S}}\bar{f}+\epsilon^{S'} f'$
& $n$ & $V_r$ & $V_\theta$ & $V_z$
& $B_r$ & $B_\theta$ & $B_z$
& $j_r$ &  $j_\theta$ & $j_z$
    \\
        \hline
         $S'$
       & $0$ & $1$   &   1     & $1$
             & $0$   &   0     & 1
             & $-1$  &  $-1$   & 0
        \\
\hline
\end{tabular}
\end{minipage}
\end{table}



As the radial coordinate is a mere parameter in the set of the reduced equations, it is convenient to replace the physical variables by the following self-similar quantities:
\begin{equation}
\tau=t,\,\,\,\ \,\,\,\eta=\frac{\zeta}{\bar{H}(r)},\,\,\,
\label{26}
\end{equation}
such that the derivatives in the new and old variables are related as follows:
\begin{equation}
\frac{\partial }{\partial t} = \frac{\partial }{\partial \tau },\,\,\,
\frac{\partial }{\partial \zeta} =\frac{1}{\bar{H}(r)} \frac{\partial }{\partial \eta }.\,\,\,
\label{27}
\end{equation}
%
Finally, a simpler form of the final equations is obtained by introducing the following scaled
dependent variables:
\begin{equation}
{\bf{v}}(\tau,r,\eta)=\frac{ { \bf{V}  }'}{\bar{c}_S (r)},\,\,\,
\nu(\tau,r,\eta)=\frac{ { n'  }}{\bar{N}(r)},\,\,\,
{\bf{b}}(\tau,r,\eta)=\frac{ { \bf{B}  }'}{\bar{B}_z (r)}. \,\,\,
\label{28}
\end{equation}
Below for convenience and with no confusion the  notation  $t$ for the time variable is reinstated instead of the new variable $\tau$.
This yields the following system of equations that depend parametrically on the radius:
\begin{equation}
\frac{1}{\bar{\Omega}(r)} \frac{\partial v_r}{\partial t}
-2v_\theta
-\frac{1}{\bar{\beta}(r)}\frac{1}{\bar{\nu}(\eta)+\nu}\frac{\partial b_r}{\partial \eta }  =
-v_z\frac{\partial v_r}{\partial \eta},
\label{29}
\end{equation}
\begin{equation}
\frac{1}{\bar{\Omega}(r)}\frac{\partial v_\theta }{\partial t}
+\frac{1}{2}v_r
-\frac{1}{\bar{\beta}(r)}\frac{1}{\bar{\nu}(\eta)+\nu}\frac{\partial b_\theta}{\partial \eta }
=-v_z\frac{\partial v_\theta}{\partial \eta},
\label{30}
\end{equation}
\begin{equation}
\frac{1}{\bar{\Omega}(r)}\frac{\partial  v_z}{\partial t}
+\frac{\bar{\nu}(\eta)}{\bar{\nu}(\eta)+\nu}
\frac{\partial}{\partial \eta }(\frac{\nu}{\bar{\nu}(\eta)})
=
-\frac{1}{2}\frac{\partial}{\partial \eta}
[v_z^2+\frac{1}{\bar{\beta}(r)}\frac{b_\theta^2+b_r^2}{\bar{\nu}(\eta)+\nu}],
\label{31}
\end{equation}
\begin{equation}
\frac{1}{\bar{\Omega}(r)}\frac{\partial \nu}{\partial t}
+\frac{\partial [(\bar{\nu}(\eta)+\nu]v_z}{\partial \eta }=0,
\label{32}
\end{equation}
\begin{equation}
\frac{1}{\bar{\Omega}(r)}\frac{\partial b_r}{\partial t}
-\frac{\partial v_r}{\partial \eta}=
-\frac{\partial (v_z b_r)}{\partial \eta },
\label{33}
\end{equation}
\begin{equation}
\frac{1}{\bar{\Omega}(r)}\frac{\partial b_\theta}{\partial t}
-\frac{\partial v_\theta}{\partial \eta}
+\frac{3}{2} b_r=
-\frac{\partial (v_z b_\theta)}{\partial \eta }
,
\label{34}
\end{equation}
supplemented by the vanishing conditions for the in-plane magnetic field and number density at infinity. Here the value of the epicyclical frequency in a Keplerian rotating medium,
$\bar{\chi}(r)=\bar{\Omega} (r)$ has been employed; $\bar{\nu} (\eta)$  is the scaled equilibrium density;
$\bar{\beta} (r)$  is the local plasma beta function:
\begin{equation}
\bar{\beta}(r)=\beta\frac{\bar{N} (r) \bar{c}_S^2 (r) }{\bar{B}_z^2 (r) },\,\,\,\,\,\,\,
\label{36}
\end{equation}
where the local parameter  $\bar{\beta}(r)$ is proportional to the characteristic plasma  beta, $\beta $.

Relations (\ref{29})-(\ref{36}) form the full nonlinear MHD problem in the thin disc approximation and
 are named as defined above, the reduced nonlinear equations.

\subsection{The linear problem}

Assuming now that the perturbations are small, the system of equations  (\ref{29})-(\ref{34}) may be linearized about the steady-state equilibrium solution. The resulting system of equations is composed of two uncoupled systems, namely which describe the Alfv\'{e}n-Coriolis (AC) and the Magnetosonic (MS) modes, respectively.

\subsection{Linear stability analysis for the Alfv\'en-Coriolis modes.}
We start by representing the perturbations up to a radius-dependent amplitude factor as follows:
\begin{equation}
f(r,\eta,t)=\hat{f}(r,\eta) \exp[-i \lambda(r)\bar{\Omega}(r)t ] ,\,\,\,
\label{43}
\end{equation}
where $\lambda(r)$ is the complex eigenvalue
\begin{equation}
\lambda=  \Lambda+i\Gamma.
\label{44}
\end{equation}

Substituting (\ref{43})-(\ref{44}) into the linearized problem results in the following system of linear ordinary differential equations for the perturbed velocity as well as in-plane magnetic field components.
That system of equations    characterizes the Alfv\'en-Coriolis waves and depends parametrically  on the radius:
\begin{equation}
-i\lambda \hat{v}_r
-2\hat{v}_\theta
-\frac{1}{\bar{\beta}(r)\bar{\nu}(\eta)}\frac{d \hat{b}_r}{d \eta }  =
0,
\label{45}
\end{equation}
\begin{equation}
-i\lambda \hat{v}_\theta
+\frac{1}{2}\hat{v}_r
-\frac{1}{\bar{\beta}(r)\bar{\nu}(\eta)}\frac{d \hat{b}_\theta}{d \eta }
=0,
\label{46}
\end{equation}
\begin{equation}
-i\lambda \hat{b}_r
-\frac{d \hat{v}_r}{d \eta}=0,
\label{47}
\end{equation}
\begin{equation}
-i\lambda \hat{b}_\theta
-\frac{d \hat{v}_\theta}{d \eta}
+\frac{3}{2} \hat{b}_r=
0.
\label{48}
\end{equation}
In addition, the Alfv\'en-Coriolis sub-system (\ref{45})-(\ref{48}) is subject to the vanishing boundary conditions for the in-plane magnetic-field components.

The linear set of equations (\ref{45})-(\ref{48}) may be reduced to the following single fourth order
ordinary differential equations for both  $\hat{b}_r$ and $\hat{b}_\theta$:
$$
\frac{d}{d\eta} \left[\frac{1}{\bar{\nu}(\eta)}
\frac{d^2}{d\eta^2} \left(\frac{1}{\bar{\nu}(\eta)}\frac{d\hat{b}_{r,\theta}}{d\eta}\right)\right]
\,\,\,\,\,\,\,\,\,\,\,\,\,\,\,\,\,\,\,\,\,\,\,\,\,\,\,\,\,\,\,\,\,\,\,\,\,\,\,\,\,\,\,\,\,\,\,\,
\,\,\,\,\,\,\,\,\,\,\,\,\,\,\,\,\,\,\,\,\,\,\,\,\,\,\,\,\,\,\,\,\,\,\,\,\,\,\,\,\,\,\,\,\,\,\,\,
\,\,\,\,\,\,\,\,\,\,\,\,\,\,\,\,\,\,\,\,\,\,\,\,\,\,
$$
\begin{equation}
+(3+2\lambda^2)\bar{\beta}(r)
\frac{d}{d\eta}  \left(\frac{1}{\bar{\nu}(\eta)}\frac{d\hat{b}_{r,\theta}}{d\eta}\right)
+ \lambda^2( \lambda^2-1)\bar{\beta}^2(r) \hat{b}_{r,\theta}=0.
\label{50}
\end{equation}
Equation (\ref{50}) is the same as the one used by \cite{Liverts and Mond 2009}   who have derived it for a model
problem under the assumption of zero radial variations of the perturbations. Here, it is important to emphasize
however that the radial coordinate is a parameter a fact that renders the radial
dependence of the perturbations arbitrary.
\cite{Liverts and Mond 2009} have solved (\ref{50}) with the aid of the Wentzel-Kramers-Brillouin (WKB) approximation for $\bar{\nu}(\eta)=\exp(-\eta^2/2)$.  Remarkably, however,   a full analytical solution of Eq. (\ref{50}) is possible for a slightly modified density profile. A detailed description of the solution may be found in \cite{Shtemler et al. 2010}. We repeat here the main results. The first and main step towards that goal is to replace the isothermal density vertical steady-state distribution $\bar{\nu}(\eta)=\exp(-\eta^2/2)$
   by the following function:
\begin{equation}
\bar{\nu}(\eta)=\mbox{sech}^2(b\eta),
\label{51}
\end{equation}
where the shape parameter $b$ is determined by the requirement that the total mass of the disc does not
change, namely:
\begin{equation}
\int^\infty_0\exp(-\eta^2/2) d\eta=
\int^\infty_0\mbox{sech}^2(b\eta) d\eta.
\label{52}
\end{equation}
The result is:
\begin{equation}
b=\sqrt{2/\pi}.
\label{53}
\end{equation}
As shown in \cite{Shtemler et al. 2010} the results derived by employing that profile are hardly distinguishable from the WKB results obtained for the true exponential distribution. In fact, notwithstanding the use of the terms true and model profiles, such a change of the equilibrium profile of the number density (that is determined by the axial momentum balance equation) may actually represent some true equilibrium that is obtained from a slightly different gravitational potential (see \cite{Spitzer 1942}, where a similar density profile has been obtained as an exact solution for flat disc-galaxies whose disc mass content is larger than the mass of the central object).


Introducing a new independent variable $\xi=\mbox{tanh}(b\eta)$, such that $-1\leq\xi\leq 1$,
a simple equation emerges, which may be cast into the following form:
\begin{equation}
(L+K^-)(L+K^+)\hat{v}_{\theta}=0,\,\,\,
\label{55}
\end{equation}
where $L$  is the Legendre operator of second order:
$$
L=\frac{d}{d\xi}[(1-\xi^2) \frac{d}{d\xi}],\,\,\,\,
K^\pm=\frac{\bar{\beta}}{2b^2}[3+2\lambda^2\pm\sqrt{9+16\lambda^2}].\,\,\,\,
\,\,\,\,\,\,\,\,\,\,\,\,\,\,\,\,\,\,\,\,\,\,\,\,\,\,\,\,\,\,\,\,\,\,\,\,\,\,\,\,\,\,\,\,\,\,\,\,\,\,\,\,\,\,
\,\,\,\,\,\,\,\,\,\,\,\,\,\,\,\,\,\,\,\,\,\,\,\,\,\,\,\,\,\,\,\,\,\,\,\,\,\,\,\,\,\,\,\,\,\,\,\,\,\,\,\,\,\,
\,\,\,\,\,\,\,\,\,\,\,\,\,\,\,\,\,\,\,\,\,\,\,\,\,\,\,\,\,\,\,\,\,\,\,\,\,\,\,\,\,\,\,\,\,\,\,\,\,\,\,\,\,\,
\,\,\,\,\,\,\,\,\,\,\,\,\,\,\,\,\,\,\,\,\,\,\,\,\,\,\,\,\,\,\,\,\,\,\,\,\,\,\,\,\,\,\,\,\,\,\,\,\,\,\,\,\,\,
\,\,\,\,\,\,\,\,\,\,\,\,\,\,\,\,\,\,\,\,\,\,\,\,\,\,\,\,\,\,\,\,\,\,\,\,\,\,\,\,\,\,\,\,\,\,\,\,\,\,\,\,\,\,
$$
Imposing now the zero boundary conditions  of $\hat{b}_\theta$  at $\eta\to \infty$ leads to the requirement that the solution of Eq.
(\ref{55}) for $\hat{v}_{\theta}$ diverges polynomially at most when $\eta\to \infty$. It is concluded therefore that $\hat{v}_{\theta}$ is proportional to the Legendre polynomials $P_k(\xi)$, and the eigenvalues $\lambda^\pm$ are now determined by the dispersion relation
\begin{equation}
K^\pm\equiv \frac{\bar{\beta}(r)}{2b^2}[3+2\lambda^2\pm\sqrt{9+16\lambda^2}]=k(k+1).
\label{57}
\end{equation}
 Setting  the arbitrary amplitude of $\hat{v}_{\theta}$  to unity, and using the linear
equations (\ref{45}) - (\ref{48}) result in the following expressions for the  eigenfunctions that are determined up to an
arbitrary radius dependent amplitude factor:
$$
\hat{v}_{r}^\pm =i\lambda^\pm_a \left(\frac{1}{2}
+\frac{3}{2}\frac{1}{\hat{\beta}}\frac{\hat{\beta} -1}{(\lambda^\pm_a)^2}\right) P_k(\xi),\,\,\
\hat{v}_{\theta}^\pm = P_k(\xi),\,\,\,
\,\,\,\,\,\,\,\,\,\,\,\,\,\,\,\,\,\,\,\,\,\,\,\,\,\,\,\,\,\,\,\,\,\,\,\,\,\,\,\,\,\,\,\,\,\,\,\,\,\,\,\,\,\,
\,\,\,\,\,\,\,\,\,\,\,\,\,\,\,\,\,\,\,\,\,\,\,\,\,\,\,\,\,\,\,\,\,\,\,\,\,\,\,\,\,\,\,\,\,\,\,\,\,\,\,\,\,\,
\,\,\,\,\,\,\,\,\,\,\,\,\,\,\,\,\,\,\,\,\,\,\,\,\,\,\,\,\,\,\,\,\,\,\,\,\,\,\,\,\,\,\,\,\,\,\,\,\,\,\,\,\,\,
\,\,\,\,\,\,\,\,\,\,\,\,\,\,\,\,\,\,\,\,\,\,\,\,\,\,\,\,\,\,\,\,\,\,\,\,\,\,\,\,\,\,\,\,\,\,\,\,\,\,\,\,\,\,
$$
$$
\hat{b}_r^\pm =\frac{k(k+1)}{2k+1}\frac{b}{6} [(\lambda^\pm_a-1) (\lambda^\pm_a+1) \hat{\beta}-3] \, [P_{k-1}(\xi) -
P_{k+1}(\xi)],
\,\,\,
\,\,\,\,\,\,\,\,\,\,\,\,\,\,\,\,\,\,\,\,\,\,\,\,\,\,\,\,\,\,\,\,\,\,\,\,\,\,\,\,\,\,\,\,\,\,\,\,\,\,\,
\,\,\,\,\,\,\,\,\,\,\,\,\,\,\,\,\,\,\,\,\,\,\,\,\,\,\,\,\,\,\,\,\,\,\,\,\,\,\,\,\,\,\,\,\,\,\,\,\,\,\,\,\,\,
\,\,\,\,\,\,\,\,\,\,\,\,\,\,\,\,\,\,\,\,\,\,\,\,\,\,\,\,\,\,\,\,\,\,\,\,\,\,\,\,\,\,\,\,\,\,\,\,\,\,\,
\,\,\,\,\,\,\,\,\,\,\,\,\,\,\,\,\,\,\,\,\,\,\,\,\,\,\,\,\,\,\,\,\,\,\,\,\,\,\,\,\,\,\,\,\,\,\,\,\,\,\,\,\,\,
$$
\begin{equation}
\hat{b}_{\theta}^\pm =\frac{k(k+1)}{2k+1}\frac{b}{4} \frac{(1-\lambda^\pm_a) (\lambda^\pm_a+1) \hat{\beta}-1 }{i\lambda^\pm_a
}\,[P_{k-1}(\xi) - P_{k+1}(\xi)],
\label{58}
\end{equation}
where $k=1,2,...$
plays the role of axial wave number, $\hat{b}_{\theta}^\pm=0$  for $\xi=\pm 1$, since $P_k(1)=1$  and $P_k(-1)=(-1)^k$   for all $k$.

Turning back to the  dispersion relation (\ref{57}), it may be written as follows:
\begin{equation}
(\lambda^\pm)^4 \hat{\beta}^2- (\lambda^\pm)^2 \hat{\beta}(\hat{\beta}+6)
+9(1-\hat{\beta})=0,\,\,\,\, \hat{\beta}=\frac{\bar{\beta} }{\bar{\beta}_{cr}^{(k)} }.
\label{59}
\end{equation}
It is thus obvious that the $k$-th mode is destabilized when the beta value crosses from bellow the threshold that is given by:
\begin{equation}
\bar{\beta}_{cr}^{(k)}=\frac{2}{3\pi}k(k+1).
\label{60}
\end{equation}
As a result, a universal (for all values of $\bar{\beta}(r)$ and  $k$) criterion for instability emerges which reads:  $\hat{\beta}(r)>1$.
Written in terms of the scaled plasma beta $\hat{\beta} $  the dispersion relation (\ref{59}) has the following
solutions for the eigenvalues of the Alfv\'en-Coriolis modes (see Fig. 1):
\begin{equation}
\lambda^\pm=\sqrt{
\frac{
\hat{\beta} +6\pm\sqrt{
     (\hat{\beta}+6)^2-36(1-\hat{\beta})
                        }
}{2\hat{\beta}}.
}
\label{61}
\end{equation}
The two eigenvalues, $\lambda^+$   and  $\lambda^-$, represent fast and slow Alfv\'en-Coriolis waves. While
the fast Alfv\'en-Coriolis modes are always stable, the number of unstable slow modes is determined by
 the plasma beta.
The eigenvalues of the slow Alfv\'en-Coriolis modes, $\lambda^-$  , are given therefore by:
$$
\lambda^-=\Lambda^-_a=\pm \sqrt{
\frac{
\hat{\beta} +6-\sqrt{
     (\hat{\beta}+6)^2-36(1-\hat{\beta})
                        }
}{2\hat{\beta}}
},     \,\mbox{Im}\{\lambda^-)=0\,\,\,\,\mbox{for}\,\,\,\hat{\beta}\leq 1,
\,\,\,\,\,\,\,\,\,\,\,\,\,\,\,\,\,\,\,\,\,\,\,\,\,\,\,\,\,\,\,\,\,\,\,\,\,\,\,\,\,\,\,\,\,\,\,\,\,\,\,\,\,\,\,\,\,\,\,\,\,\,\,\,\,\,\,\,\,\,\,\,\,\,\
\,\,\,\,\,\,\,\,\,\,\,\,\,\,\,\,\,\,\,\,\,\,\,\,\,\,\,\,\,\,\,\,\,\,\,\,\,\,\,\,\,\,\,\,\,\,\,\,\,\,\,\,\,\,\,\,\,\,\,\,\,\,\,\,\,\,\,\,\,\,\,\,\,\,\
$$
\begin{equation}
\lambda^-=i\Gamma^-_a=\pm i \sqrt{
\frac{
\hat{\sqrt{
     (\hat{\beta}+6)^2+36(\hat{\beta}-1)-\beta}_z -6
                        }
}{2\hat{\beta}}
},     \,\mbox{Re}\{\lambda^-)=0\,\,\,\,\mbox{for}\,\,\,\hat{\beta}> 1.
\label{62}
\end{equation}
The eigenvalues are imaginary, and the system is spectrally unstable if $\hat{\beta} >1$, and real for
$\hat{\beta} <1$  in which case the system is stable. In particular, the minimal critical unscaled plasma
beta that is needed for instability is determined by the first unstable slow Alfv\'en-Coriolis mode, $k=1$, and is given by $\bar{\beta}_{cr}^{(1)}=0.42$.
The unstable modes are the well known MRIs  and from Eq. (\ref{60}) it is easy to
calculate how many of them are excited for a given value of the plasma beta.
 Thus, there are $k$ unstable modes for $\bar{\beta}_{cr}^{(1)}\leq\bar{\beta}(r)\leq\bar{\beta}_{cr}^{(k+1)}$.
 In particular, Eq. (\ref{60}) yields approximately the square-root law for the number of the unstable modes as a function of the plasma beta:
\begin{equation}
k<\sqrt{3\pi\bar{\beta}(r)/2}.
\label{63}
\end{equation}
Relation (\ref{60}) or its simplified version (\ref{63}) is significant for the consequent modeling of non-linear development of the instability. It is finally emphasized that the stability criterion as well as the number of unstable modes depend on the radius. Thus, different areas within the disc may be characterized by different stability properties as well as different number of unstable modes.

The family of the fast Alfv\'en-Coriolis modes, $\lambda^+(\hat{\beta})$, is characterized by frequencies that are
much larger than the Keplerian frequency for small values of the scaled plasma beta (large values of axial wave number or
small plasma beta) and tends to the Keplerian value at large plasma beta values:
\begin{equation}
\lambda^+=\Lambda^+_a=\pm \sqrt{
\frac{
\hat{\beta} +6+\sqrt{
     (\hat{\beta}+6)^2-36(1-\hat{\beta})
                        }
}{2\hat{\beta}}
},     \,\mbox{Im}\{\lambda^+)=0.
\label{64}
\end{equation}

Expressed in terms of the scaled plasma beta, a single figure depicts all possible stable as well as unstable
modes. This is shown in Fig. 1. The maximal growth rate for the unstable modes is achieved around
$\hat{\beta} \approx3$, which for a given plasma beta value determines the  axial wave number of the
fastest growing modes.


\begin{figure*}
\includegraphics[width=120mm]{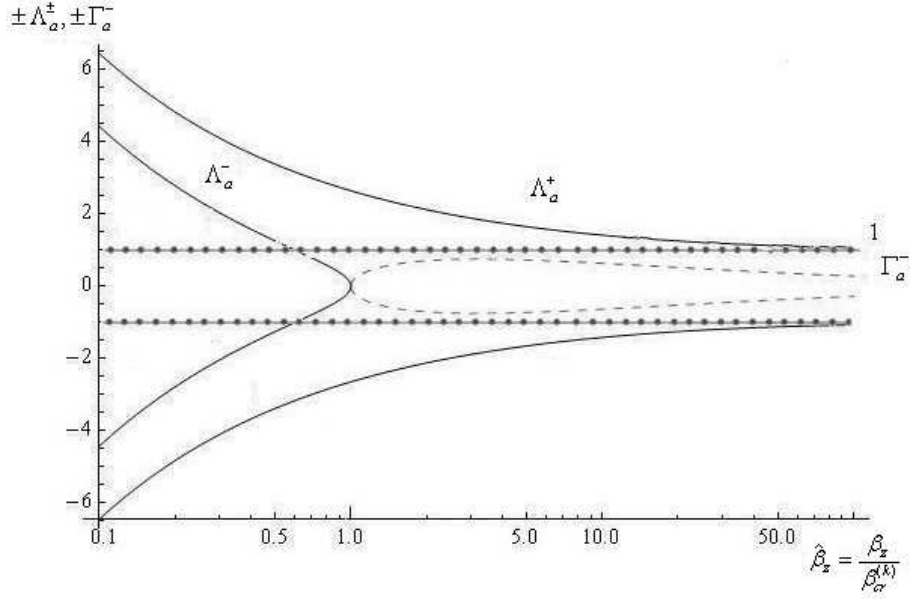}
  \caption{Growth rates $\pm\Gamma^-_a$  (dashed lines) for the unstable Alfv\'en-Coriolis (MRI) modes and frequencies
  $\pm\Lambda^\pm_a$  (solid lines) for the Alfv\'en-Coriolis oscillations vs universal scaled plasma beta,
  $\hat{\beta} =\bar{\beta}(r)/ \bar{\beta}_{cr}^{(k)}$, $\bar{\beta}_{cr}^{(k)}=2k(k+1)/(3\pi)$ for the model number density
  $\bar{\nu}=\mbox{sech}^2(\sqrt{2/\pi}\eta)$;
$k=1,2,\dots$  is the axial wave number. Meshed straight-line asymptotes at $\hat{\beta} \gg 1$  are the scaled
Keplerian frequencies, $\pm\Lambda^\pm_a=\pm 1$.
  }
\label{Fig. 2}
\end{figure*}

An illustration of the perturbed toroidal magnetic field is presented in Fig. 2. The perturbations are indeed localized within the effective height of the disc which weakly depends on the axial wave number. This corresponds to a finite distance between the turning points -- the natural characteristics of the problem solution in the WKB approximation (see \cite{Liverts and Mond 2009}).

In order to compare the current results to well known results for infinite homogeneous cylinders \cite{Balbus and Hawley 1991} the growth rate of the unstable modes is depicted in Fig. 3 as a function of the axial wave number  $K$, where
 \begin{equation}
K=k \frac{L_a}{ H\sqrt{2}}\equiv \frac{k}{\sqrt{2\bar{\beta}}},
\label{65}
\end{equation}
$L_a=V_a/\Omega$ is the Alfv\'en length scale, $z=H\sqrt{2}$ is the effective height of the diffused disc at which the equilibrium number density, $\bar{n}\sim\exp[-z^2/(2H^2)]$, falls by factor $e^{-1}$.
 The effective  wave number $K$ is the discrete thin-disc analog of the continuous wave number for  infinite cylindrical discs. For fixed value of the local plasma beta,  $\bar{\beta}=0.41,\,\,0.5,\,\,1.5,\,\,2.5,\,500$,  the discrete set of the points in the plane $\{K,\Gamma^{-}_a\}$ is presented by one of the interpolating curves $1, 2, 3, 4, 5$, respectively. The number of the discrete  points on each interpolating curves  corresponds to the admissible values of the Alfv\'en-Coriolis  mode number $k=1,2,3,...$ for which $\bar{\beta}_{cr}^{(1)}\leq\bar{\beta}_{cr}^{(k)}\leq\bar{\beta}(r)$.
 Also, the range of unstable $k$-values is widening as the value of $\bar{\beta}$  is increased.
 In particular, at large plasma beta the corresponding set of  the points due to their large number (curve 5) should tend to the continuous curve for infinite cylinder geometry in \cite{Balbus and Hawley 1991}. For finite plasma beta there is a discrete number of points on each curves in Fig. 3, where the left bound corresponds to the  first Alfv\'en-Coriolis mode, $k=1$.

\begin{figure*}
\includegraphics[width=175mm]{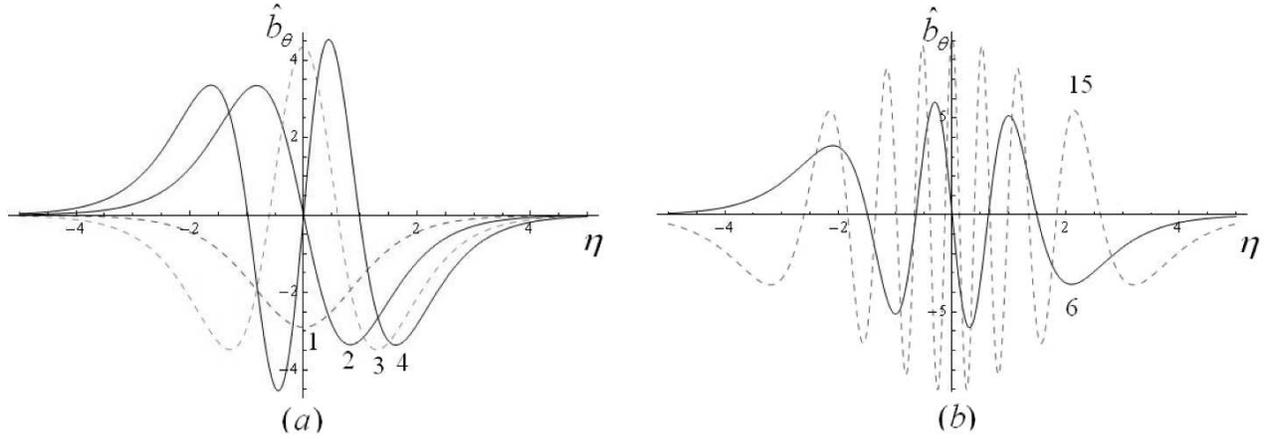}
 \caption{
 The toroidal component of the perturbed magnetic field  $\hat{b}_\theta$ vs self-similar axial variable $\eta=z/H(r)$  for the Legendre polynomials with the axial wave numbers (a)  $k=1, 2, 3,4$ and (b)   $k=6,\,15$.
All curves are calculated for the fixed value of  $\hat{\beta}_z=1.5$ ($\hat{\beta}_z=\bar{\beta}_z/\bar{\beta}_{cr}^{(k)}$ ,  $\bar{\beta}_{cr}^{(k)}=b^2k(k+1)/3$, dashed and solid  curves correspond to the odd  and even  $k$, respectively).
  }
\label{Fig. 4}
\end{figure*}
 Thus, for beta values close to  $\bar{\beta}_{cr}^{(1)}$, the number of unstable modes is small and the disk stability properties significantly deviate from those predicted by the infinite cylinder model.

\begin{figure*}
\includegraphics[width=115mm]{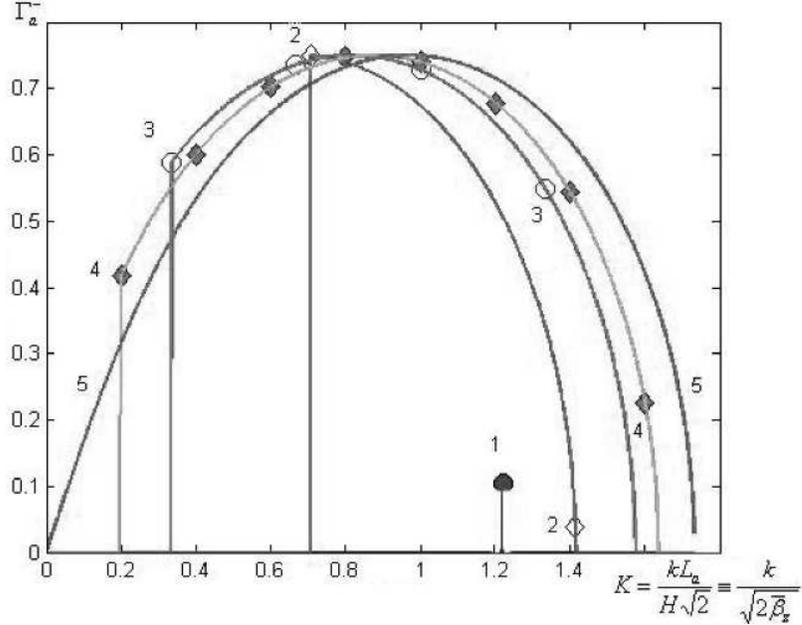}
 \caption{
Growth rates $\Gamma^-_a$   for the slow Alfv\'en-Coriolis modes vs effective wave number $K=\frac{k L_a}{H \sqrt{2}}\equiv\frac{k}{\sqrt{2\bar{\beta}}}$,
 $k=1,2,3,...$  is the number of the Alfv\'en-Coriolis modes, calculated for the model number density  $\bar{\nu}=\mbox{sech}(b\eta)$. Interpolating curves $1, 2, 3, 4, 5$ correspond to
 $\bar{\beta}=0.41,\,\,0.71,\,\,1.5,\,\,2.5,\,500$, respectively.
  }
\label{Fig. 4}
\end{figure*}

\subsection{Linear stability problem for the magnetosonic modes}
The magnetosonic sub-system of equations is analyzed now under the same model density profile that was employed in the previous section, namely $\bar{\nu}(\eta)=\mbox{sech}^2(b\eta)$. This unifies the treatment of all linear modes in the system and sets the base for further nonlinear analysis. Thus, assuming the form (\ref{44}) for the perturbed variable, expressed in terms of the new independent variable, $\xi=\mbox{tanh}(b\eta)$, the magnetosonic system [i.e., the lineraized version of eqs. (\ref{31}) and (\ref{32})] may be reduced to the following single ordinary differential equation for the perturbed number density:

\begin{equation}
(1-\xi ^2)\frac{\partial ^2 \hat{\nu}}{\partial \xi ^2}+[\frac{\lambda ^2}{1-\xi ^2} +2] \hat{\nu}=0,
\label{66}
\end{equation}
subject to the boundary conditions $\nu (\pm 1)=0$.
Transforming to a new dependent variable,

\begin{equation}
\hat{\nu} (\xi )=\sqrt{1-\xi ^2 }f(\xi),
\label{67}
\end{equation}
the solution to eq. (\ref{66}) is given by:
 \begin{equation}
\hat{\nu}(\xi )=\sqrt{1-\xi ^2 }[C_1 f_1(\xi )+C_2 f_2(\xi )],
\label{68a}
\end{equation}
where,
\begin{eqnarray}
f_1&=&\bigl (\frac{1-\xi }{1+\xi } \bigr )^{-\mu /2}(\mu -\xi )\\
f_2&=&\bigl (\frac{1-\xi }{1+\xi } \bigr )^{\mu /2}(\mu +\xi ),
\label{68}
\end{eqnarray}
and $\mu = \sqrt{1-\lambda ^2 }$.
Solutions that satisfy the boundary conditions exist only for $\lambda ^2 >0$. This means that the magnetosonic modes are stable. Furthermore, the magnetosonic spectrum that corresponds to the model density profile is continuous. The two independent solutions for $\lambda =2$ are depicted in Fig. 4 where it is easy to see that the arbitrary (positive) value of $\lambda ^2$ actually determines the axial wave number.

\begin{figure*}
\includegraphics[width=155mm]{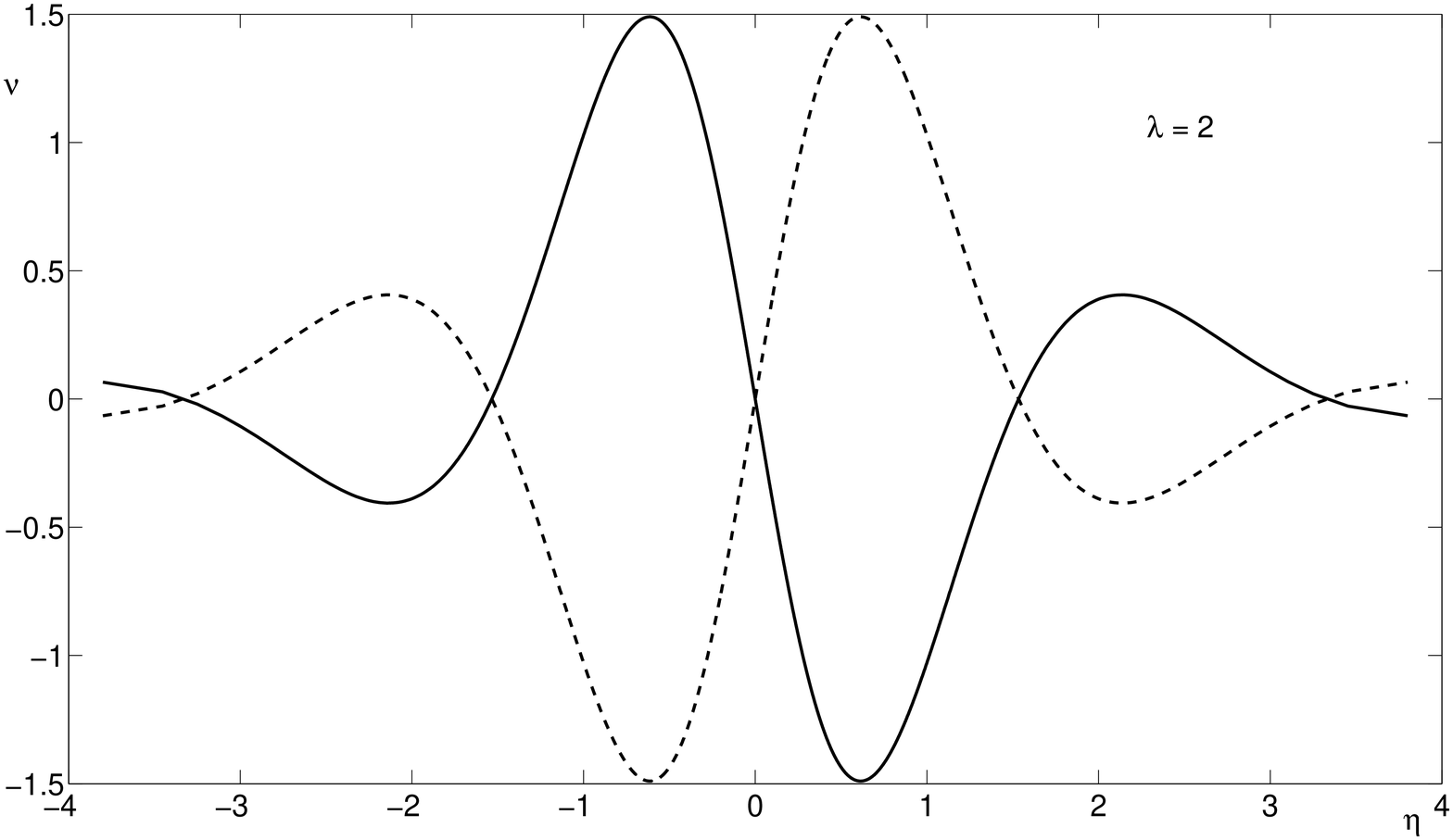}
 \caption{
Two independent solutions of eq. (\ref{66}) for $\lambda =2$. The full (broken) line corresponds to the solution with
$f_1$ ($f_2$).
  }
\label{Fig. 4}
\end{figure*}

\section{Weakly nonlinear analysis of the MRI}
The viability of the MRI as a generator of turbulence in thin discs, and consequently as an important cause of angular
momentum transport, depends of course on its nonlinear dynamical development and saturation mechanisms. In order to gain
insight and  to provide guidelines to full nonlinear numerical simulations we carry out in this section a weakly nonlinear analysis of the interaction of the unstable MRI with the stable magnetosonic modes.
\subsection{A piece-wise constant model}
The eigenvalues as well as the eigenfunctions of the linearized MHD thin-disk system of equations, serve as building blocks of the weakly nonlinear analysis of the unstable modes, namely the MRI. As was shown in the previous section the eigenfunctions may be expressed in terms of the Legendre as well as some well defined hypergeometric functions. However, in order to simplify the vast amount of algebra that usually accompanies any attempt to deviate from linearity, a simpler piece-wise constant model of the steady state is introduced. As will be shown, the resulting spectrum is very close to that obtained from the diffused profiles (see eq. (\ref{51})).

We start by introducing the following piece-wise constant axial profile for the number density:
\begin{equation}
\bar{\nu} (\eta ) = \cases{ \nu _0\;\;\;\; | \eta | \leq 1  \cr  0 \;\;\;\;\;\; | \eta |>1,\cr}
\label{pw}
\end{equation}

where $\nu _0 =\sqrt{\pi/2}$.
\subsubsection{The spectrum of the AC modes}

Inserting expression (\ref{pw}) into eqs. (\ref{45})-(\ref{48}), results after some algebra in the following dispersion relation for the AC modes:
\begin{equation}
\bigl [\pi ^2 (k+\frac{1}{2})^2-2\lambda ^2 \tilde{\beta}\bigr ]\bigl [\pi ^2 (k+\frac{1}{2})^2-(3+\lambda ^2)\tilde{\beta}\bigr ]-4\lambda ^2\tilde{\beta}^2=0,\;\;\; k=0,1,2...,
\label{69}
\end{equation}
where $\tilde{\beta}=\nu _0\bar{\beta}$, and the eigenfunctions are given by:
\begin{eqnarray}
\Bigl [\hat{b}_r(\eta ), \hat{b}_{\theta}(\eta )\Bigr ]&=&(\hat{b}_{r0}, \hat{b}_{\theta 0})\cos[\pi(k+\frac{1}{2})\eta ]\\
\Bigl [ \hat{v}_r(\eta ), \hat{v}_{\theta}(\eta )\Bigr ]&=&(\hat{v}_{r0}, \hat{v}_{\theta 0}
)\sin[\pi(k+\frac{1}{2})\eta ]
\end{eqnarray}
As in the diffused case discussed above, also for the piece-wise constant profiles the number $K$ of unstable modes (i.e. with $\Gamma _k ^2 = -\lambda _k^2 >0$ for $k\leq K$) depends on the value of $\tilde{\beta}$. For $K=1$ the threshold for the instability is $\tilde{\beta}_{c}=\pi ^2/12$. The latter corresponds to $0.65$ for the diffused profiles (i.e. after eq. (\ref{62})) which is close to the exact value.
\subsubsection{The spectrum of the magnetosonic modes}
Expression (\ref{pw}) is inserted now into the linearized version of eqs. (\ref{31}) and (\ref{32}). Using representation (\ref{43}) for the perturbations, and utilizing the fact that the perturbed axial velocity is an antisymmetric function about the midplane, results in the following dispersion relation for the magnetosonic modes:
\begin{equation}
\lambda _k  ^2=\pi ^2(k+\frac{1}{2})^2,\;\;\;k=0,1,2...,
\label{70}
\end{equation}
and the following eigenfunctions:
\begin{eqnarray}
\hat{v}_z=a_k\sin[\pi(k+\frac{1}{2})\eta]&\\
\hat{\nu}=\nu _0a_k\cos[\pi(k+\frac{1}{2})\eta]&.
\label{71}
\end{eqnarray}

\subsubsection{Weakly nonlinear analysis}
We consider now a beta value which is slightly above $\tilde{\beta}_c$. In that case there is only one MRI mode ($k=0$), which is characterized by a small growth rate $\Gamma$. The relation between $\Gamma$ and $\tilde{\beta}$ may be inferred from the dispersion equation. (\ref{69}). The result may be written in the following way:
\begin{equation}
\tilde{\beta}=\frac{\pi ^2}{12}(1+\alpha ^2 \Gamma ^2),
\label{72}
\end{equation}
where $\alpha ^2=7/9$. Thus, up to a radius-dependent multiplicative factor, the eigenfunctions of the first unstable mode ($k=0$) are given by:
\begin{eqnarray}
b_r(\eta ,t)&=&a(t)\bigl (\frac{\pi}{3}+\frac{4\pi}{27}\Gamma ^2 \bigr)\cos\frac{\pi}{2}\eta\\
b_{\theta}(\eta ,t)&=&-a(t)\frac{2\pi}{9}\Gamma \cos\frac{\pi}{2}\eta ,
\label{73}
\end{eqnarray}
where $a(t)=e^{\gamma \bar{\Omega}(r)t}$ is the initially small amplitude of the MRI. Similar expressions may be written for $v_r(\eta ,t)$ and $v_{\theta}(\eta ,t)$. Returning now to eq. (\ref{31}) and (\ref{32}), we see that the MRI mode nonlinearly excites a magnetosonic wave. To lowest order in $\Gamma$ and $a$, the forced part of the latter is described, therefore, by:
\begin{eqnarray}
\rho _f &=&-\frac{1}{3}a^2\bigl ( \cos\pi \eta +1 \bigr )\\
 v_{z,f}&=&\frac{2}{3}a^2 \Gamma \bigl ( \frac{ \sin\pi \eta}{\pi} +\eta \bigr )
 \label{74}.
 \end{eqnarray}
Obviously the excited magnetosonic waves feed back on the MRI.

The aim of the weakly nonlinear analysis is to find an ordinary differential equation that describes the time evolution of the amplitude $a(t)$ of the single unstable MRI mode. In order to do that it is realized first that the transition to instability (when $\beta$ reaches the value $\beta _c$ from below) occurs when the linearized system has a double zero eigenvalue. This situation is known as the Takens-Bogdanov bifurcation \cite{Guckenheimer and Holmes 1983}. As a result the sought after equation is expected to be of second order as opposed to first order equations (like the Landau-Ginzburg one) that characterize systems that bifurcate through a simple zero eigenvalue. Thus following Arter \cite{Arter 2009} it is conjectured that the typical Takens-Bogdanov amplitude equation for ideal MHD is of the form:
\begin{equation}
\frac{d^2a}{dt^2}=\Gamma ^2 a-\alpha a^3.
\label{bt}
\end{equation}
In order to calculate $\alpha$ it is noticed that the equilibrium solution of eq. (\ref{bt}) provides an asymptotic approximation of the steady state solutions of eqs. (\ref{29}-\ref{34}). The latter may be obtained by employing the Poincar\'{e}-Lindstedt method. The latter yields the following expression for the steady state amplitude of the radial component of the perturbed magnetic field:
\begin{equation}
b_r ^0=\frac{\pi \sqrt{14}}{3\sqrt{3}}\Gamma
\label{br}
\end{equation}
Equating the right hand side of eq. (\ref{bt}) to zero and inserting expression (\ref{br}) results in the following expression for $\alpha$:
\begin{equation}
\alpha = \frac{27}{14\pi ^2}.
\label{alfa}
\end{equation}
All is ready now to examine the solutions of eq. (\ref{bt}). As an example consider the case $\Gamma =0.01$, $a(0)=0.001$, and $\dot{a}(0)=\Gamma a(0)$. The results of the numerical solution of eq. (\ref{bt}) for that particular case are shown in Fig. \ref{Fig. 4}. For small values of $a(t)$ it behaves according to the linearized equation, however as the nonlinear term kicks in it varies on a much faster time scale, thus giving it a bursty appearance. The saturation level is about one order of magnitude higher than the initial condition. In order to examine the viability of the weakly nonlinear analysis the results presented in Fig. \ref{Fig. 4} are compared to the results obtained form the numerical solution of the fully nonlinear set of reduced equations (\ref{29}-\ref{34}). The initial conditions are given by eqs. (\ref{73}) and (\ref{74}) while $\beta$ is determined from eq. (\ref{72}) with $\Gamma =0.01$. That solution for the time evolution of the amplitude of the radial component of the perturbed magnetic field is displayed in Fig. \ref{Fig. 5}. It is clear that the weakly nonlinear analysis provides an accurate description of the evolution of the MRI near the threshold. Numerical solutions for initial conditions that are far from threshold indicate that the nature of oscillatory saturation of the instability is preserved. In that case however, the saturation occurs on realistic time scale of the order of ten rotation times.

\begin{figure*}
\includegraphics[width=155mm,height=80mm]{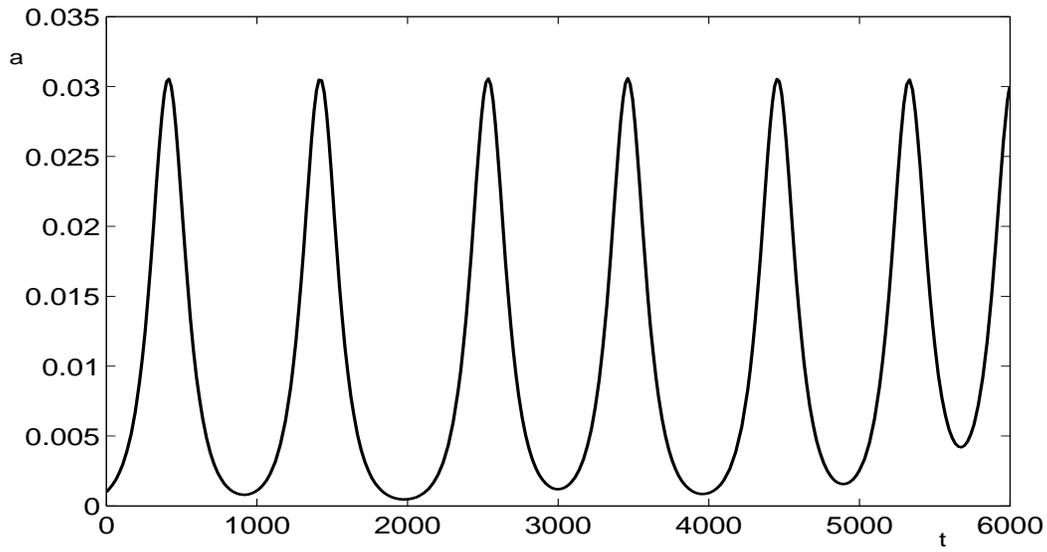}
 \caption{
Solution of eq. (\ref{bt}) for $\Gamma = 0.01$, $a(0)=0.001$, $\dot{a}(0)=\Gamma a(0)$.
  }
\label{Fig. 4}
\end{figure*}

\begin{figure*}
\includegraphics[width=155mm,height=80mm]{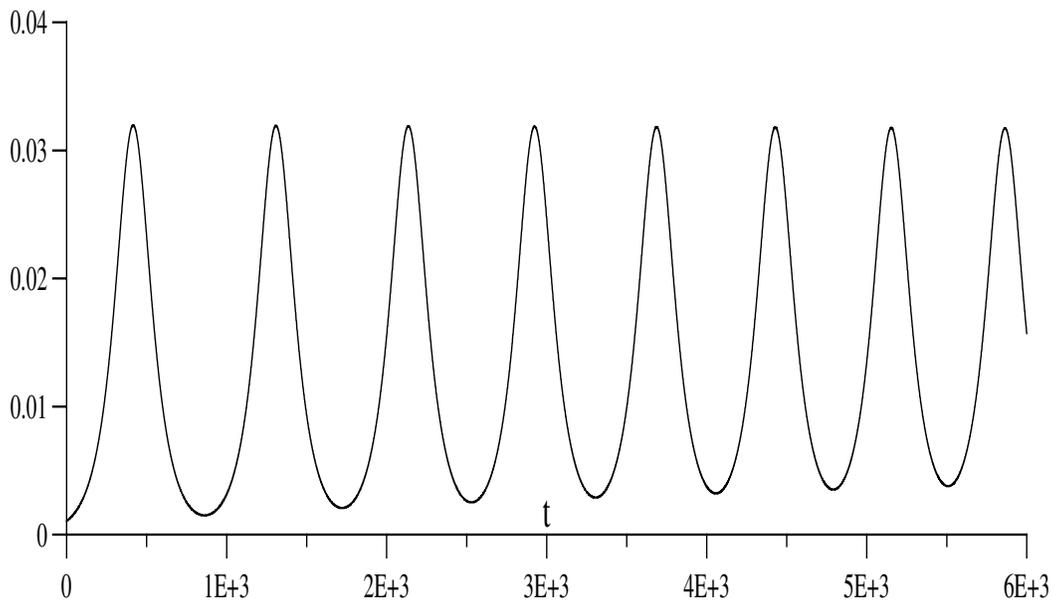}
 \caption{
The evolution of the amplitude of the radial component of the perturbed magnetic field as obtained from the numerical solution of eqs. (\ref{29}-\ref{34}) for the parameters that appear in Fig. \ref{Fig. 4}.
  }
\label{Fig. 5}
\end{figure*}

\section{Conclusions}
The full spectrum of the MHD modes in thin rotating axially-isothermal discs under the influence of axial magnetic field has been obtained analytically. The number of unstable MRI modes as a function of the plasma beta has been derived, as well as the dispersion relation for the stable magnetosonic waves. A weakly nonlinear analysis points out one possible mechanism of saturation of the MRI, namely energy transfer from the latter to the stable magnetosonic modes. For the current model this occurs as oscillatory saturation due to the conservative nature of the Ideal MHD equations. Adding enough dissipation, the degeneracy of the bifurcation point may be removed, which ultimately leads to a first order Landau-Ginzburg like equation for the amplitude, instead of the second order equation given in eq. (\ref{bt}). Saturation of the latter type has been investigated by Umurhan et al. \cite{Umurhan et al 2007}.


\begin{theacknowledgments}
  The current work was supported by grant no. 180/10 of the Israeli Science Foundation.\\
  The authors are grateful to O.M. Umurhan for his enlightening and crucial comments about the weakly nonlinear analysis, and to V. Borisov for carrying out the calculations for Fig. \ref{Fig. 5}.
\end{theacknowledgments}



\bibliographystyle{aipproc}   

\bibliography{sample}

\IfFileExists{\jobname.bbl}{}
 {\typeout{}
  \typeout{******************************************}
  \typeout{** Please run "bibtex \jobname" to optain}
  \typeout{** the bibliography and then re-run LaTeX}
  \typeout{** twice to fix the references!}
  \typeout{******************************************}
  \typeout{}
 }



\end{document}